# Perspective on tailoring quantum coherence with electron beams

Nahid Talebi*[1]

*Institute for Experimental and Applied Physics, Kiel University, 24118 Kiel, Germany*

***Abstract*** – Examining and controlling the interaction between semiconductor quantum qubits and their environment can boost semiconductor quantum technologies, which have many applications in table-top quantum computing hardware. Electron beams in electron microscopes have opened up a new avenue for the quantum-coherent probing of semiconductor excitations and strong-coupling effects. Here, I provide a brief overview of recent advancements in electron-beam probes for investigating quantum coherence in semiconductors and two-dimensional materials, complemented by my perspective on using electron beams to manipulate the entanglement and correlations between quantum systems.

## 1. Introduction and Motivation

Over the past decade, remarkable progress has been achieved in resolving excitonic, plasmonic, and polaritonic dynamics in semiconductors and two-dimensional materials with unprecedented spatial resolution using electron beams [1-6]. Recently, the interaction of free-electrons with optical nanostructures has evolved from a tool for nano-spectroscopy toward a platform for exploring quantum coherence [7-13]. In this *Perspective*, I discuss how some of these advances pave the way for using electron beams not only as passive quantum-coherent probes but also as active agents for shaping quantum states, generating entanglement and correlations in condensed-matter systems.

The interconnection between quantum optics, electron microscopy, and solid-state quantum information is expected to enable new approaches for probing and controlling qubits using both light and free-electron wave packets. Electron-beam-based quantum probing is an emerging paradigm, making this perspective relevant for both optics and quantum technology communities.

## 2. Electron Beams as Coherent Probes

Probing coherent dynamics of quantum systems such as quasi-two-level systems requires a coherent pump, so that its interaction with the quantum system generates a coherent superposition of its individual states, but also probes that enable direct read-out of that coherence. In this context, two individual approaches have been pursued. The first relies on shaping the electron wave packet, that is, turning the beam into a coherent superposition of different energy states. This approach produces a frequency comb that allows for a coherent read-out of the system's state [14-16].

The second employs interactions between electron beams, an electron-driven photon source (EDPHS), and the quantum system, forming a Ramsey-type interferometric scheme [17-20]. In this configuration, the radiation emitted by the EDPHS and the electron beam sequentially interact with the sample [21] (Fig. 1a). The optical pulse from the EDPHS creates a coherent superposition of the quantum states—comprising both electronic and vibronic degrees of freedom—of a nitrogen-vacancy defect as a qubit in a thin film of hexagonal boron nitride. The subsequent interaction of the electron beam with this superposition generates a cathodoluminescence (CL) signal that exhibits quantum-coherent, Ramsey-like interference fringes. The visibility of these fringes depends sensitively on the temporal delay between the EDPHS excitation and the arrival of the electron beam at the sample. For delays exceeding the coherence time of the defect excitation, the interference visibility completely vanishes (Fig. 1b) [13]. Theoretically, a formulation based on the reduced density matrix of the emitters, treating electron beams as an incoherent pump, captures the evolution of the Ramsey fringes and the decoherence effect [11, 22].

[1] Corresponding author: talebi@physik.uni-kiel.de

The efficiency of the Ramsey interferometry scheme is influenced by several experimental factors. First, the CL signal generated by the EDPHS must be sufficiently strong to induce a measurable, albeit potentially non-optimal, coherent superposition in the sample. Second, efficient delivery of the EDPHS radiation to the sample is required; this has been demonstrated in Ref. [11] through the design of EDPHS structures generating collimated beams. Third, the collection efficiency of the CL signal must support a sufficiently high signal-to-noise ratio (SNR), which can be achieved using optimized collection optics such as parabolic mirrors, yielding SNR values on the order of ~20. Current experiments are primarily performed using continuous electron beams in scanning electron microscopes. However, the role of the electron wave packet duration in probing quantum coherence remains an open question. Recent theoretical studies suggest that electron wave packets can effectively probe quantum-coherent excitations of prepared superposition states, provided that the wave packet duration exceeds $T = 2\pi\omega^{-1}$, where $\omega$ is the transition frequency of the qubit [23]. The fact that the Ramsey fringes are experimentally observed, demonstrates that the continuous electron beams of a Schottky emitter provides the necessary temporal duration.

The strength of cathodoluminescence (CL) spectroscopy lies in its ability to resolve the spectral features of individual solid-state qubits and quantum emitters with nanometer spatial resolution, providing an unprecedented tool for exploring the photophysics of different classes of single emitters [24-26]. Interestingly, electron beams can also be used to generate emitters in wide-bandgap materials under suitable excitation conditions and beam parameters [27]. The EDPHS-based Ramsey interferometry scheme discussed above further extends these capabilities by enabling direct measurements of the decoherence timescales of individual emitters.

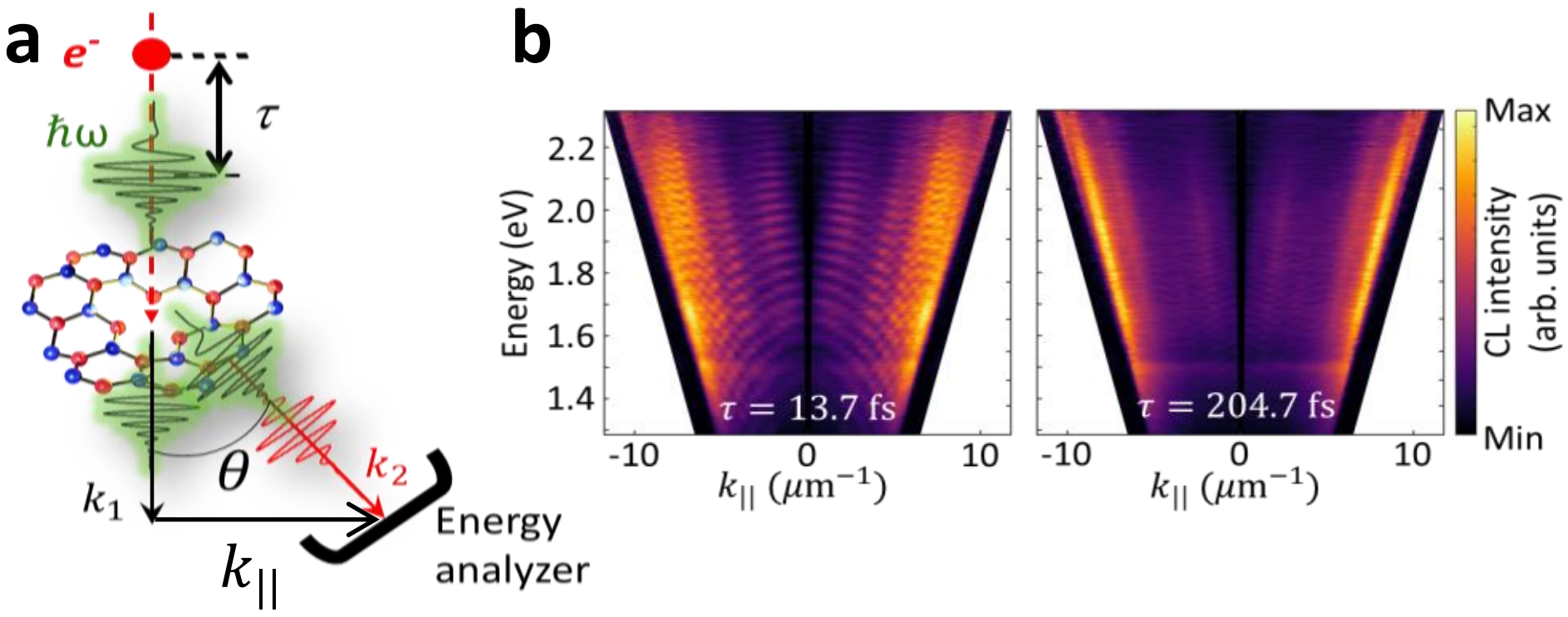

**Fig. 1.** (a) Ramsey interferometry by the sequential interaction of the EDPHS radiation (green pulse) and the moving electron with a single defect. (b) Ramsey interference fringes of the CL signal versus the photon energy and lateral wavenumber acquired at depicted delays between the EDPHS radiation and the electron beam [13].

Beyond mapping decoherence pathways, engineered electron–light–matter interactions could enable the tailored control of quantum information in solid-state platforms, bridging concepts from quantum optics and electron microscopy. Such an engineered interaction mechanism could be enabled by combining coherent read-out of qubits using structured electron beams combined with an EDPHS.

## 3. Toward Quantum Control and Entanglement

More ambitiously, the interaction mechanisms could be extended toward two-dimensional electronic spectroscopy, by integrating an EDPHS capable of generating two sequential optical pulses (Fig. 2a). This could be achieved for example by combining Smith-Purcell radiation sources with adaptable mirrors, which reflects generated Smith-Purcell radiation towards the sample. Smith–Purcell emission is generated when an electron beam moves parallel to a periodic grating, where the grating provides the momentum required for phase-matched coupling of the electron field to radiative diffraction modes [28, 29].

Two-dimensional electronic spectroscopy has proven to be a powerful technique for examining the coupling strength between vibronic and electronic states. Merging this method with shaped electron wave packets as coherent probes of quantum dynamics would yield a unique tool for performing two-dimensional electronic spectroscopy on single solid-state qubits with nanometer spatial resolution. Several mechanisms could be used to generate such shaped electrons, including stimulated Compton scattering [30], inelastic scattering of electrons from structured light fields [31], or interactions with plasmonic near-field distributions [32].

Sequential interactions of electron beams with two qubits (with the states $|g\rangle_1, |g\rangle_2, |e\rangle_1, |e\rangle_2$, with g and e representing ground and excited) coupled to a photonic resonator with photon-number states $|n\rangle$) can be tailored to generate entanglement between qubits This goes beyond the so-called *parametric electron–field interaction* [33], which typically produces entanglement only between electrons and photons (Fig. 2b). Within the weak interaction scheme, a moving electron takes a quantum walk on its energy ladder denoted by $|U_e - l\hbar\omega\rangle$, with $l$ being an integer. We assume here that the transition frequency of the two-level system and the photonic resonator frequencies are perfectly tuned. The electron interacting with the first qubit generate a superposition state in the form of $c_1|U_e, g_1, g_2, l=0\rangle + c_2|U_e - \hbar\omega, e_1, g_2, l=0\rangle + c_3|U_e - \hbar\omega, g_1, g_2, l=1\rangle$, with $c_i$ being complex-valued amplitudes satisfying the normalization condition $\sum_i |c_i|^2 = 1$. The second interaction leads to the wavefunction

$$\begin{aligned}|\psi\rangle = {} & \alpha_1|U_e, g_1, g_2, 0\rangle + \\ & \alpha_2|U_e - \hbar\omega, g_1, e_2, 0\rangle + \alpha_3|U_e - \hbar\omega, e_1, g_2, 0\rangle + \alpha_4|U_e - \hbar\omega, g_1, g_2, 1\rangle + \\ & \alpha_5|U_e - 2\hbar\omega, e_1, e_2, 0\rangle + \alpha_6|U_e - 2\hbar\omega, e_1, g_2, 1\rangle + \alpha_7|U_e - 2\hbar\omega, g_1, e_2, 1\rangle + \alpha_8|U_e - 2\hbar\omega, g_1, g_2, 2\rangle,\end{aligned} \tag{1}$$

with $\sum_i |\alpha_i|^2 = 1$. By projecting the wavefunction in eq. (1) to electron energy subspaces, either $|U_e - \hbar\omega, 0\rangle\langle U_e - \hbar\omega, 0|$ or $|U_e - 2\hbar\omega, 1\rangle\langle U_e - 2\hbar\omega, 1|$, i.e., by detecting electrons within selected energy ranges the entangled state, the state $\beta_1|g_1, e_2\rangle + \beta_2|e_1, g_2\rangle$ is obtained, with $\sum_i |\beta_i|^2 = 1$. This

projection-based measurement thus provides a pathway toward heralded entanglement generation between qubits mediated by free electrons.

The proposed experimental configuration has slight similarities with geometries that have been proposed and theoretically explored with optical beams propagating in waveguides attached to emitters [34]. However, the inclusion of the electron beam introduces a significant advantage: the electron–matter interaction can be precisely tuned by controlling the impact parameter. This tunability enables the transition from weak coupling to strong correlation regimes and offers a heralded scheme for generating entanglement among multiple emitters [10].

Efficient probing of quantum coherence and entanglement generation further requires careful consideration of the electron wave packet duration. Specifically, the wave packet duration should exceed the inverse transition frequency, while remaining shorter than the electron travel time between the two qubits to preserve their distinguishability. For field-emission electron sources with typical kinetic energies of ~30 keV and energy spreads of ~0.2 eV, the corresponding longitudinal and temporal broadenings are on the order of $\delta L = 324\,\mathrm{nm}$ and $\delta t = 3.3\,\mathrm{fs}$, respectively. Several classes of visible-range emitters, including vacancy centers in diamond or hexagonal boron nitride, as well as semiconductor quantum dots (e.g., CdS, CdSe, GaAs), are suitable candidates for such experiments, provided that their spatial separation exceeds $\delta t = 50\,\mathrm{fs}$ ($\delta L = 5\,\mu\mathrm{m}$). For photoemission-based electron sources, the longitudinal coherence length is significantly larger, requiring correspondingly larger qubit separations.

Since the decoherence times of these emitters typically range from hundreds of femtoseconds to nanoseconds, the electron–qubits interaction time remains well below the intrinsic decoherence timescale. Therefore, solid-state decoherence does not fundamentally limit the feasibility of the proposed schemes, which substantiates as well the wave-function-based derivations in eq. (1). Moreover, the photon statistics of electron-beam excited emitters is also dependent on electron-beam statistics [26]. Therefore, a careful use of electron parameters to achieve single-photon emissions for entanglement measurements is necessary.

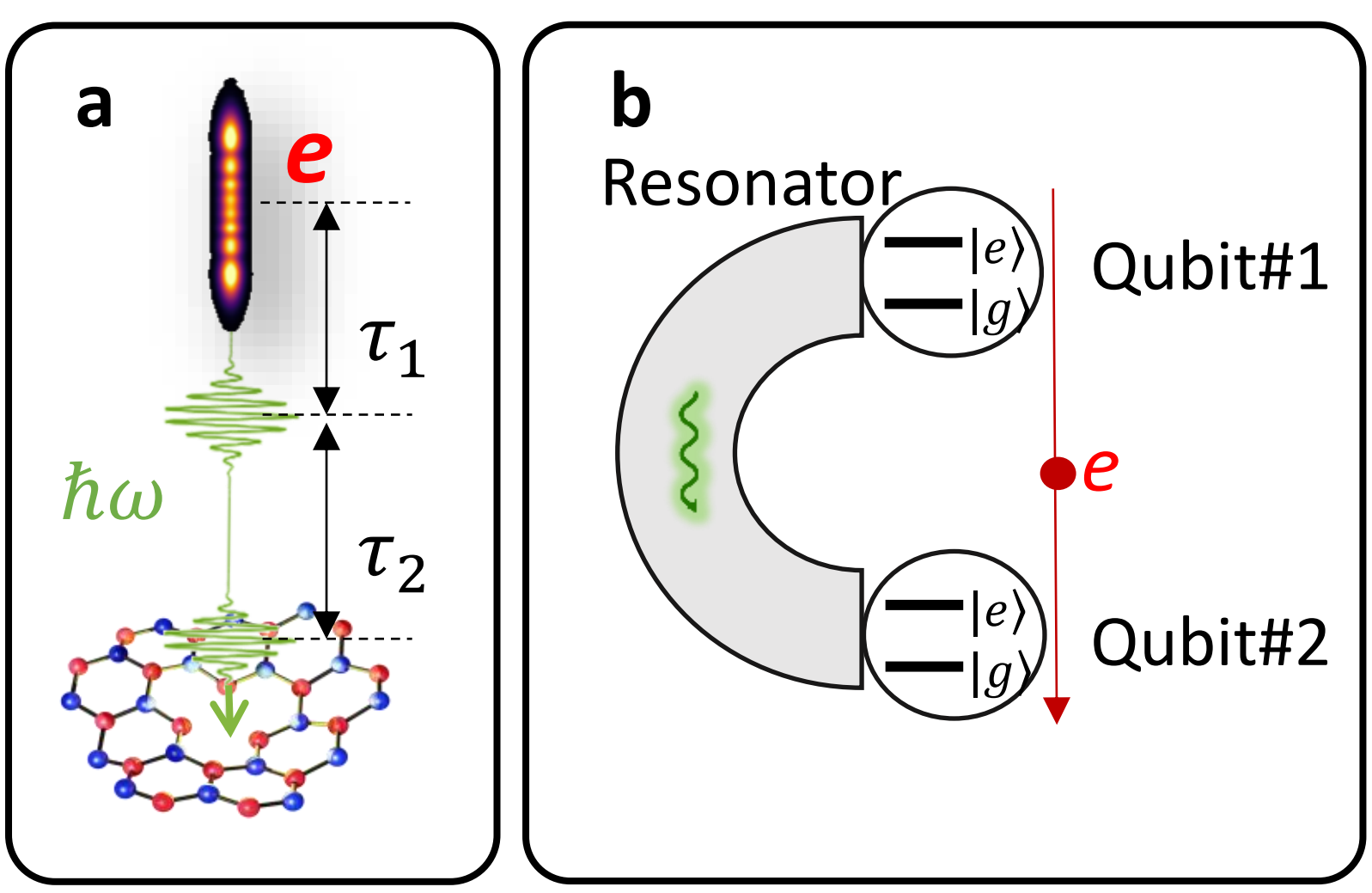

**Fig. 2.** (a) A scheme for merging shaped electron beams with two sequential optical pulses generated from EDPHS, for performing two-dimensional electronic spectroscopy of single solid-state qubits. (b) Generating entanglement between qubits via coherent interaction of an electron beam with a resonator combined with two qubits.

As a summary, ultrafast all-optical and near-field methods—such as femtosecond pump–probe spectroscopy combined with ultrafast scattering-type scanning near-field optical microscopy—have the potential to achieve nanometer spatial resolution together with femtosecond temporal resolution. However, direct schemes for coherent control and readout of solid-state qubits based on these approaches remain limited.

Electron-beam-based techniques inherently provide broadband excitation and nanometer spatial localization without the need for external optical focusing. In particular, electron-driven photon sources enable ultrafast optical excitation directly within the electron microscope, while CL detection provides spectroscopic readout with nanometer resolution. Moreover, the electron–matter interaction strength and spatial selectivity can be tuned via the electron trajectory and impact parameter, offering a degree of control not readily accessible in purely optical or scanning probe approaches. These features position electron-beam-based methods as a complementary platform for probing and controlling quantum coherence in solid-state systems.

## 4. Future Directions

The growing interest in exploiting electron microscopes for quantum-sensitive measurements offers exciting possibilities—not only for exploring material systems with enhanced sensitivity but also for probing and tailoring quantum correlations in condensed-matter platforms. These advanced characterization techniques open new pathways toward a deeper understanding of quasiparticle interactions and correlations in materials.

Among the approaches discussed here, sequential electron–light–matter interaction schemes based on EDPHSs are already compatible with electron microscopy infrastructures and offer direct access to nanoscale optical excitation and spectroscopic readout. With continued advances in optimized collection optics, and nanofabricated photonic architectures, proof-of-principle demonstrations of coherent control and electron-mediated entanglement between emitters appear feasible within the coming years. Advanced configurations for EDPHS and photonic cavities based on three-dimensional nanoprinting [35], offer a viable route for intense and directional sources as well as vacuum-friendly optical collection schemes [36].

In the longer term, combining these approaches with tailored photonic or plasmonic resonators [37] may enable scalable control of interactions between multiple qubits. Electron-beam shaping may enable a scheme for electron-induced superposition generation and model-selective readouts of quantum emitters coupled to cavities [38-41].

Near-term progress toward electron-beam-based qubit control will likely be assessed through intermediate performance metrics rather than full gate fidelities. Relevant benchmarks include selective excitation probability, coherence preservation during interaction, Ramsey-fringe visibility, readout contrast, and the repeatability of sequential electron-mediated protocols. Successful optimization of these quantities in single-emitter experiments would provide a clear route toward future fidelity tests of electron-mediated one- and two-qubit operations.

Altogether, experimental schemes outlined here allow for exploring quantum-coherent dynamics in qubit networks in solid-state systems and mapping their couplings with phonons, guided waves, and excitons. Tailoring quasi-particle interactions in hybrid systems of emitters coupled to phonons or excitons provide a pathway for generating optically-tuned new phases of materials in hybrid systems. Therefore, advanced characterization techniques as suggested here lead to new metrology schemes for quantum-sensitive measurements. This could open pathways toward hybrid quantum materials and new metrology schemes based on quantum-sensitive measurements with electron microscopes.

**Acknowledgement**

I acknowledge funding from the Volkswagen Foundation (Momentum Grant), European Research Council (ERC) under the European Union's Horizon 2020 research and innovation program under grant agreement no. 101017720 (EBEAM), from the European Research Council (ERC Consolidator Grant UltraSpecT with no. 101170341; ERC Proof-of-Concept Grant UltraCoherentCL with no. 101157312) and from Deutsche Forschungsgemeinschaft.